# The orbital and superhump periods of the deeply eclipsing dwarf nova SDSS J122740.83+513925.9

Jeremy Shears, Steve Brady, Jerry Foote, Donn Starkey & Tonny Vanmunster

**Abstract**

During June 2007 the first confirmed superoutburst of the eclipsing dwarf nova SDSS J122740.83+513925.9 was observed using CCD photometry. The outburst amplitude was at least 4.7 magnitudes. The orbital period was measured as 0.06296(5) d from times of the 31 observed eclipses. Time series photometry also revealed superhumps with a period of 0.0653(3) d, thereby establishing it to be a UGSU-type system. The superhump period excess was 3.7% and the maximum peak-to-peak amplitude of the superhumps was 0.35 magnitudes. The eclipse duration declined from a maximum of 23 min at the peak of the outburst to about 12 mins towards the end. The depth of the eclipses was correlated with the beat period between the orbital and superhump periods.

**Introduction**

Dwarf novae are semi-detached binary stars in which a cool main sequence secondary star loses mass to a white dwarf primary. Material is transferred from the secondary via Roche-lobe overflow, but because it carries substantial angular momentum, does not settle on the primary immediately, instead forming an accretion disc. As material builds up in the disc, thermal instability drives the disc into a hotter, brighter state causing an outburst in which the star brightens by several magnitudes. Dwarf novae of the SU UMa family (UGSU) also exhibit *superoutbursts* which last several times longer than normal outbursts and may be up to a magnitude brighter. During a superoutburst the light curve of a UGSU star is characterized by superhumps. These are modulations which are a few percent longer than the orbital period and are thought to be caused by precession of the accretion disc. For a full account of dwarf novae the reader is directed to reference 1.

The dwarf nova SDSS J122740.83+513925.9 (hereafter SDSS1227) was first identified spectroscopically in the Sloan Digital Sky Survey (SDSS [2]) database, having a g* magnitude (similar to visual) of 19.10 [3]. A deep doubling of its Balmer emission lines suggested that it was a high inclination system with the likelihood of eclipses, but no follow-up photometry or spectroscopy was conducted at the time. The star is located in Canes Venatici at 12h 27 min 40.83 sec +51 deg 39 min 25.9 sec (J2000).

**June 2007 Outburst**

The outburst of SDSS1227 discussed in this paper was first detected by Patrick Schmeer on 2007 June 4.106 at magnitude 15 [4]. To our knowledge, this is the first outburst of this star to be detected. Schmeer relayed the observation to TV, who confirmed it on June 4.888. TV also reported two eclipses and the presence of superhumps, showing it to be a member of the UGSU family [4]. An image of the field is shown in Figure 1.

The authors conducted photometry of SDSS1227 using the instrumentation shown in Table 1 and according to the observation log in Table 2. Julian Dates are used in this paper in truncated form where JD = JD – 2454000. All observations were unfiltered (C, clear filter), except for those made with the Liverpool Telescope, where a Johnson V filter was used. Images were dark-subtracted and flat-fielded prior to being measured using differential photometry relative to the comparison star GSC3458-372, whose V magnitude was found by Henden to be 15.254 [5]; the same study also listed SDSS1227 at 19.285V.

The overall lightcurve of the outburst is shown in Figure 2. The apparent scatter in the magnitudes is of course due to presence of the eclipses and the superhumps. SDSS1227 was observed to be at its brightest on June 4.94 at 14.5C. Considering the average magnitude outside eclipses, the brightness showed an approximately linear decline at 0.15 mag/d from JD 256 to 270, followed by a sharper decline to about magnitude 18.8. Assuming quiescence to be near 19.3, as measured by Henden [5], the outburst amplitude was at least 4.8 magnitudes. It is impossible to determine the duration of the outburst precisely since we do not know how long the star had been in outburst before discovery by Schmeer. Furthermore, there are rather few data points defining the approach to quiescence. However, from the data we do have it is clear that the outburst lasted at least 15 days. It is unlikely that any significant rebrightening occurred since we imaged the field of SDSS1227 on 8 nights between 287 and 296, but found it to be undetectable, with an upper detection limit of mag 17.8.

**Measurement of the orbital period**

Photometry from the time series runs is shown in Figure 3, all plotted at the same magnitude and time scale. This shows recurrent eclipses superimposed on an underlying superhump modulation, each of which will be considered in more detail later.

A total of 31 eclipses were observed during the outburst. The times of minimum of each eclipse were determined according to the Kwee & van Woerden method [6] using the *Peranso* software [7]. These are listed in Table 3, along with errors given by the Kwee & van Woerden method, where the eclipses are labelled with the corresponding orbit number starting from 0. The orbital period was then calculated by a linear fit to these times of minima as $P_{orb}$ = 0.06296(5) or 90m 39.7+/- 4.3s. The eclipse time of minimum ephemeris is:

$$JD = 2454256.4110 + 0.06296(5)*E$$

The O-C (Observed – Calculated) residuals of the eclipse minima relative to this ephemeris are shown in Figure 4. This suggests that the period remained constant during the period of observations. However, we note that there is considerable scatter in O-C values which is probably due to the difficulty in isolating eclipse minima relative to other large-scale changes in the light curve in the form of the underlying superhumps and the fact that each eclipse was defined by rather few data points.

**Measurement of superhump period**

Prominent superhumps were displayed throughout the outburst. Peak-to-peak amplitude was ~0.3 mag on the first night (JD 256), increasing to 0.35 mag by JD 258 and then declining gradually until the final night of time series photometry (JD 264) when it was 0.2 mag. However, analysis of the superhumps was complicated by the presence of the eclipses, which often made the humps difficult to determine. In most cases this confounded attempts to determine the times of maxima of individual humps. Instead, we carried out period analysis using the DCDFT algorithm in *Peranso*, having removed the data during the eclipses.

Analysis of the time-series data from JD 258 (June 7; Figure 3(b)), which shows 4 superhumps, resulted in the power spectrum shown in Figure 5. The strongest signal arises from the superhumps and reveals a superhump period $P_{sh}$ = 0.0647(24) d (several other algorithms including Lomb-Scargle, CLEANest and ANOVA yielded the same period). The uncertainty is calculated using the Schwarzenberg-Czerny method [8]. Removing $P_{sh}$ from the power spectrum leaves only weak signals, none of which have significant relationship to the superhump or orbital periods.

We also carried out a similar period analysis on the data from JD 263 – 264 (June 12-13), the last time series runs conducted during the outburst. The power spectrum yielded a superhump period $P_{sh}$ = 0.0651(11) d. Removing $P_{sh}$ from the power spectrum left no other significant signals.

The two values of $P_{sh}$ are consistent with each other, given their respective errors, and provide no evidence for a significant period change during this part of the outburst. Therefore we performed period analysis on the combined data from JD 258 and JD 263-264, after subtracting the mean and linear trend from the light curves, which gave $P_{sh}$ = 0.0653(3) d. The data folded on this value of $P_{sh}$ are shown in the phase diagram in Figure 6. This suggests that the phase was coherent throughout the period, so we adopt the superhump period as 0.0653(3) d. This $P_{sh}$ value represents a period excess, ε, of 3.7%. Such value of ε is in line with the distribution of period excess vs orbital period of UGSU systems given in Figure 6.6 of Reference 1.

We also note that the light curve on the first night of photometry (JD 256; Figure 3(a)) appears to show smaller humps between the superhumps, although it is difficult to be certain as these smaller humps almost coincide with eclipses. Period analysis of this data set was less than satisfactory as the run was rather short and by the time the eclipses had been removed, rather less data was available for analysis. The power spectrum contained a very broad peak centred on 0.0654 d, which we assume is $P_{sh}$, and a smaller but sharper peak at 0.0308(14) d, which is close to half the value of $P_{orb}$. There is also a suggestion of a small hump or inflexion on the rising side of the superhumps on the next night of photometry (JD 258; Figure 3(b)). These features may be similar to the outburst orbital humps seen early in the 1997 outburst of DV UMa [9].

**The nature of the eclipses**

One of the most interesting aspects of SDSS1227 are the deep eclipses. Figure 7 shows that the eclipse duration was greatest at the peak of the outburst (23 min) followed by an approximately linear decline as the outburst progressed, with the final eclipses being about half the duration of the first ones observed. This is a common feature of eclipses during dwarf nova outbursts and is due to the accretion disc being largest at the peak of the outburst. [9]. It would be interesting to examine the duration of eclipses during quiescence, when the accretion disc is expected to be at a minimum, but such a project requires a large telescope due to the faint quiescent magnitude of SDSS1227.

As noted above, a cursory examination of the time series light curves presented in Figure 3 shows that the eclipse depth is strongly affected by the location of the superhump: eclipses are shallower when hump maximum coincides with eclipse. This is commonly observed in eclipsing UGSU systems including DV UMa and IY UMa [9, 10]. The effect is demonstrated in Figure 8 where the eclipse depth is plotted against beat phase of orbit and superhump (two beat cycles shown for clarity). Zero beat phase is the phase at which hump maximum coincides with eclipse. We calculated the beat period (i.e. the precessional period of the elliptical accretion disc), $P_{beat}$, as 1.757 d from the following equation obtained from reference 1:

$$P_{beat} / P_{sh} = P_{beat} / P_{orb} - 1$$

using our values $P_{sh}$ = 0.0653 d and $P_{orb}$ = 0.06296 d obtained above.

Eclipse profiles offer great diagnostic power in probing the components of dwarf novae. It is sometimes possible to distinguish features such as the structure of the accretion disc, including the location and size of the hot spot where the material from the secondary star impacts the disc and the position of the white dwarf. However, such analysis of fine structure in eclipse profiles generally requires a higher time resolution than achieved in our observations and is ideally carried out at quiescence when the accretion disc is smallest. This would be another project worthy of a large telescope in the future. Nevertheless, we produced an eclipse profile by combining the 5 eclipses between JD 254.56 and 254.88 (from the time series photometry shown in Figure 3(h); these eclipses were selected as they show minimal interference from underlying superhumps). This average profile (Figure 9) shows that the eclipse is smooth and almost symmetrical, although there is a suggestion that ingress is slightly sharper than egress. Such a profile has been seen in other dwarf novae during decline from outburst, including IP Peg [11, 12], which may be due to a contribution from the hot spot during egress.

**Conclusion**

Our observations of the first confirmed outburst of SDSS J122740.83+513925.9 show that it is a deeply eclipsing UGSU-type dwarf nova. The outburst lasted at least 15 days and the amplitude was at least 4.8 magnitudes. It showed an approximately linear decline of 0.15 mag/d for the first 14 days of the outburst, followed by a sharper decline towards quiescence. No rebrightening was observed. Analysis of the eclipse times yielded an orbital period of 0.06296(5) d. Time-series photometry also revealed common superhumps with a period of 0.0653(3) d and a 3.7% superhump period excess. The maximum peak-to-peak amplitude of the superhumps was 0.35 magnitudes. There was no migration to late superhumps. The eclipse duration declined from a maximum of 23 min at the peak of the outburst to approximately half this value later in the outburst, indicating a shrinking accretion disc. The depth of the eclipses was correlated with the beat period between the orbital and superhump periods.

Since eclipsing dwarf nova offer interesting possibilities to probe the structure of such systems via eclipse profiling, high speed photometry of SDSS1227 at quiescence and during future outbursts would be an interesting project for a large telescope. We also encourage observers, both visual and CCD-equipped, to monitor this star for future outbursts with the aim of establishing the outburst frequency and supercycle length.

**Acknowledgements**

The authors thank Prof. Joe Patterson, Columbia University, New York, for his support and encouragement and for publicising the outburst of SDSS1227 via the Center for Backyard Astrophysics (CBA), an informal global network of observers dedicated to cataclysmic variables which he coordinates. We gratefully acknowledge the use of the Bradford Robotic Telescope operated by the Department of Cybernetics, University of Bradford, located on Tenerife. We were fortunate to be able to conduct photometry with the Liverpool Telescope on La Palma, operated by Liverpool John Moores University, thanks to time being made available to the BAA and we record our gratitude to Nick James for facilitating the scheduling of these observations. Finally we thank the referees for constructive comments which have improved the paper.

**Addresses:**
JS: "Pemberton", School Lane, Bunbury, Tarporley, Cheshire, CW6 9NR, UK [bunburyobservatory@hotmail.com]
SB: 5 Melba Drive, Hudson, NH 03051, USA [sbrady10@verizon.net]
JF: Center for Backyard Astrophysics (Utah), 4175 E. Red Cliffs Drive, Kanab, UT 84741, USA [jfoote@scopecraft.com]
DS: Center for Backyard Astrophysics (Indiana), 2507 County Road 60, Auburn, IN 46706, USA [starkey73@mchsi.com]
TV: Center for Backyard Astrophysics (Belgium), Walhostraat 1A, B-3401 Landen, Belgium [tonny.vanmunster@cbabelgium.com]

| Observer | Telescope | CCD | Filter |
|---|---|---|---|
| JS | 0.28 m SCT <br> 0.35 m SCT (Bradford Robotic Telescope)[a] <br> 2 m reflector (Liverpool Telescope)[a] | Starlight Xpress SXV-M7 <br> FLI MaxCam ME2 + EEV CCD47-10 chip <br> EEV CCD42-20 chip | C <br> C <br> V |
| SB | 0.4 m reflector | SBIG ST-8XME | C |
| JF | 0.6 m reflector | SBIG ST-8XE | C |
| DS | 0.4 m reflector | SBIG ST-10MXE | C |
| TV | 0.35 m SCT | SBIG ST-7MXE | C |

**Table 1: Instrumentation**
[a] these telescopes were used for discrete data rather than for time series photometry

| Date (2007) | Start time (JD-2454000) | Duration (hours) | No. of images | No. of eclipses | Observer |
|---|---|---|---|---|---|
| June 4  | 256.388 | 3.1 | 160 | 2 | TV |
| June 7  | 258.579 | 4.6 | 270 | 3 | SB |
| June 7  | 258.658 | 4.4 | 250 | 3 | JF |
| June 8  | 260.413 | 3.9 | 293 | 3 | JS |
| June 9  | 261.410 | 3.3 | 307 | 2 | JS |
| June 10 | 261.587 | 5.9 | 197 | 4 | DS |
| June 10 | 261.700 | 3.2 | 193 | 2 | JF |
| June 12 | 263.589 | 6.7 | 228 | 4 | DS |
| June 12 | 264.412 | 2.1 | 215 | 1 | JS |
| June 13 | 264.590 | 6.6 | 219 | 5 | DS |
| June 13 | 264.664 | 4.1 | 106 | 2 | JF |
| June 15 | 266.660 | 0.0 | 3   | 0 | SB |
| June 16 | 267.585 | 0.0 | 5   | 0 | SB |
| June 19 | 270.585 | 0.0 | 2   | 0 | SB |
| June 20 | 271.564 | 0.0 | 1   | 0 | SB |
| June 27 | 278.585 | 0.0 | 1   | 0 | SB |
| June 28 | 280.465 | 0.0 | 2   | 0 | JS |
| July 2  | 283.553 | 0.0 | 1   | 0 | JS |
| July 3  | 284.628 | 0.0 | 2   | 0 | SB |

**Table 2: Log of observations**
(only positive detections of SDSS1227 are included in the log)

| Orbit no. | Eclipse time of minimum (JD −2454000) | Error (d) |
|---|---|---|
| 0 | 256.4110 | 0.0001 |
| 1 | 256.4742 | 0.0009 |
| 35 | 258.6157 | 0.0001 |
| 36 | 258.6769 | 0.0001 |
| 36 | 258.6771 | 0.0002 |
| 37 | 258.7389 | 0.0001 |
| 37 | 258.7395 | 0.0001 |
| 38 | 258.8033 | 0.0002 |
| 64 | 260.4405 | 0.0001 |
| 65 | 260.5032 | 0.0001 |
| 66 | 260.5668 | 0.0001 |
| 80 | 261.4485 | 0.0001 |
| 81 | 261.5120 | 0.0001 |
| 83 | 261.6360 | 0.0001 |
| 84 | 261.6995 | 0.0002 |
| 85 | 261.7630 | 0.0002 |
| 85 | 261.7632 | 0.0003 |
| 86 | 261.8251 | 0.0002 |
| 86 | 261.8256 | 0.0001 |
| 115 | 263.6514 | 0.0007 |
| 116 | 263.7138 | 0.0004 |
| 117 | 263.7779 | 0.0004 |
| 118 | 263.8410 | 0.0005 |
| 128 | 264.4690 | 0.0001 |
| 130 | 264.5956 | 0.0001 |
| 131 | 264.6587 | 0.0003 |
| 132 | 264.7219 | 0.0004 |
| 132 | 264.7220 | 0.0002 |
| 133 | 264.7842 | 0.0001 |
| 133 | 264.7852 | 0.0003 |
| 134 | 264.8476 | 0.0009 |

**Table 3: Eclipse times of minimum**

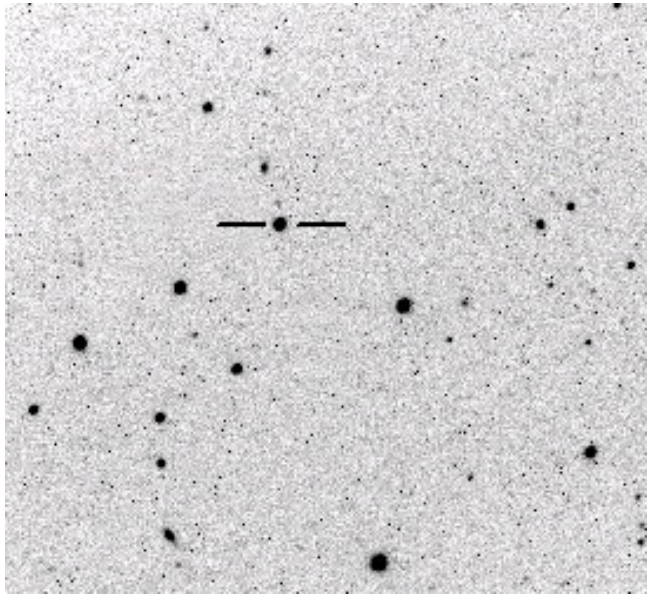

**Figure 1: SDSS1227 in outburst**
Mag. 15.2C. Field 7 x 7 arcmin, N at top, E to left (Jerry Foote)

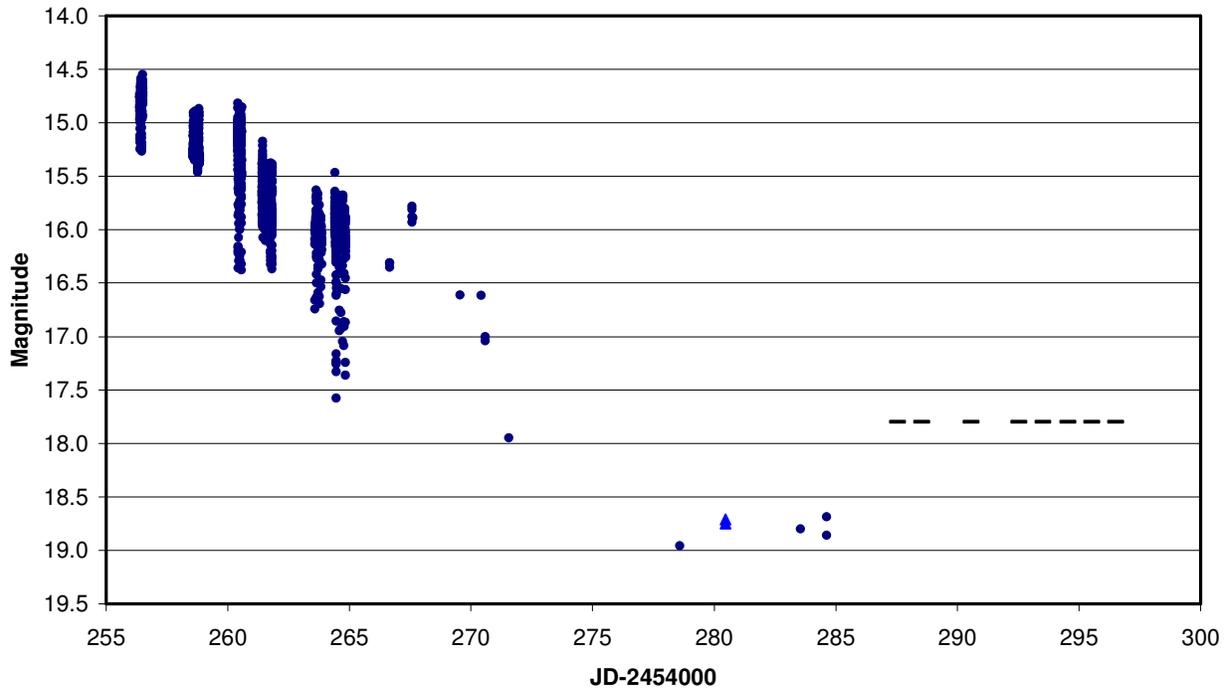

**Figure 2: Light curve of the outburst**
Unfiltered CCD photometry, except for two V-filter measurements represented by triangles. Observations represented by a dash are upper limits (i.e. <17.8) established from observations made with the Bradford Robotic Telescope.

(a) 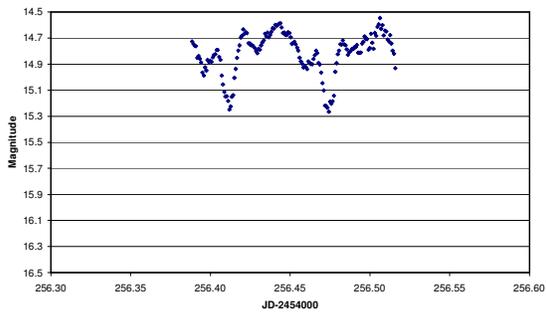
(b) 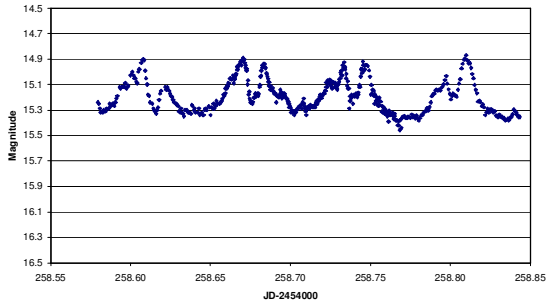
(c) 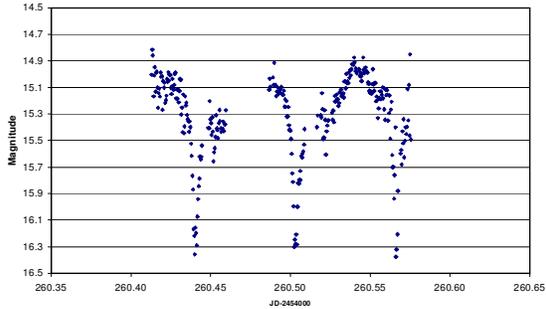
(d) 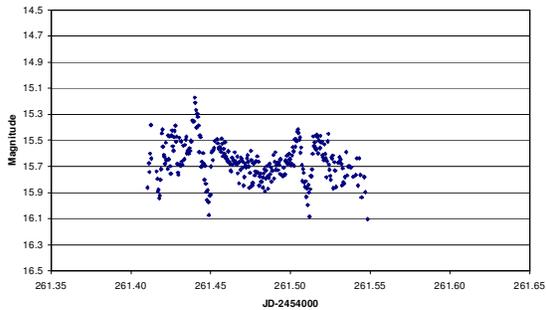
(e) 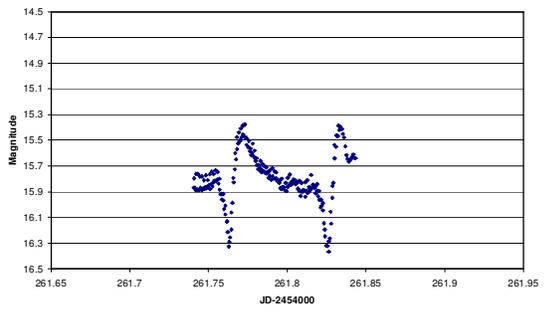
(f) 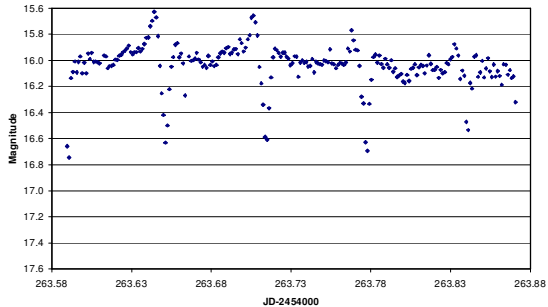
(g) 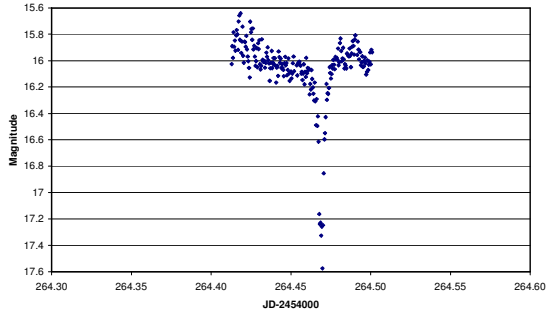
(h) 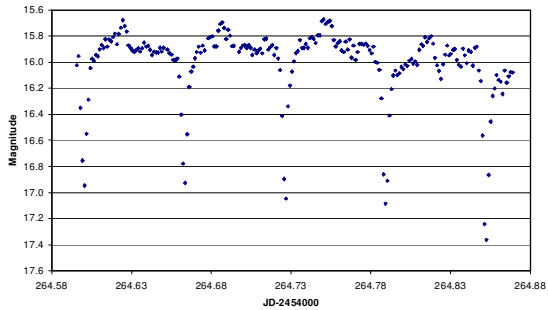

**Figure 3: Time series photometry of SDSS1227**

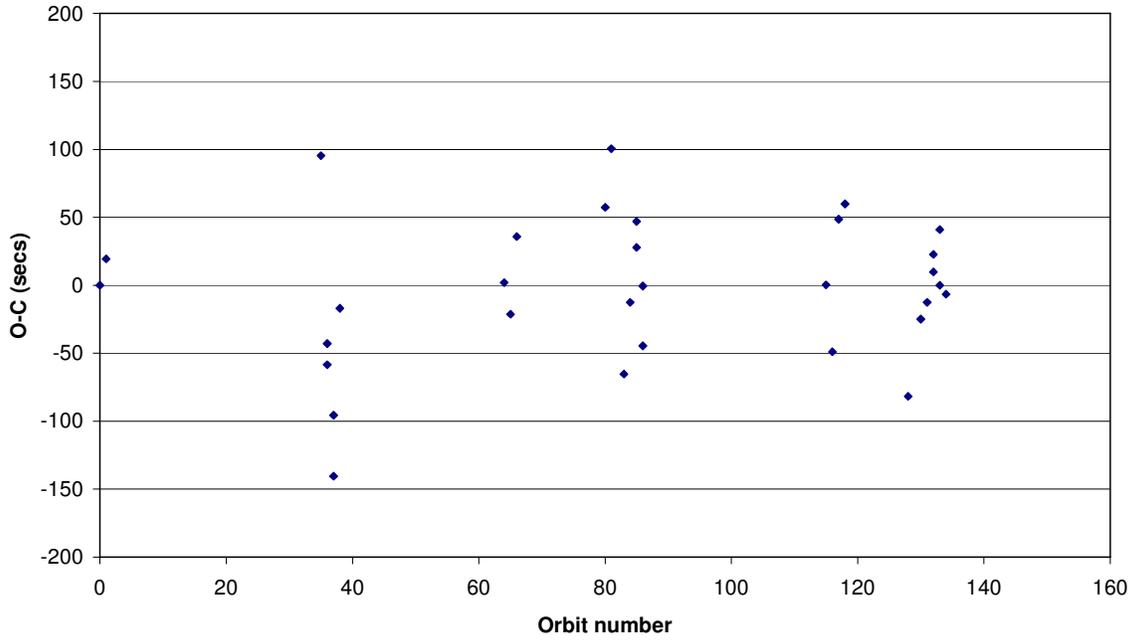

**Figure 4: O-C residuals for the eclipses**

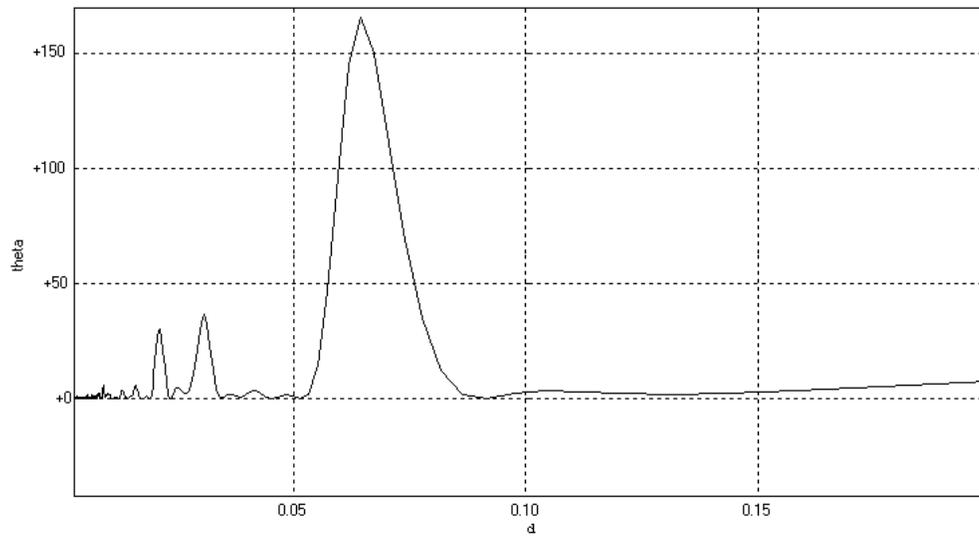

**Figure 5: Power spectrum of data from JD 258 (June 7) after removal of eclipses**

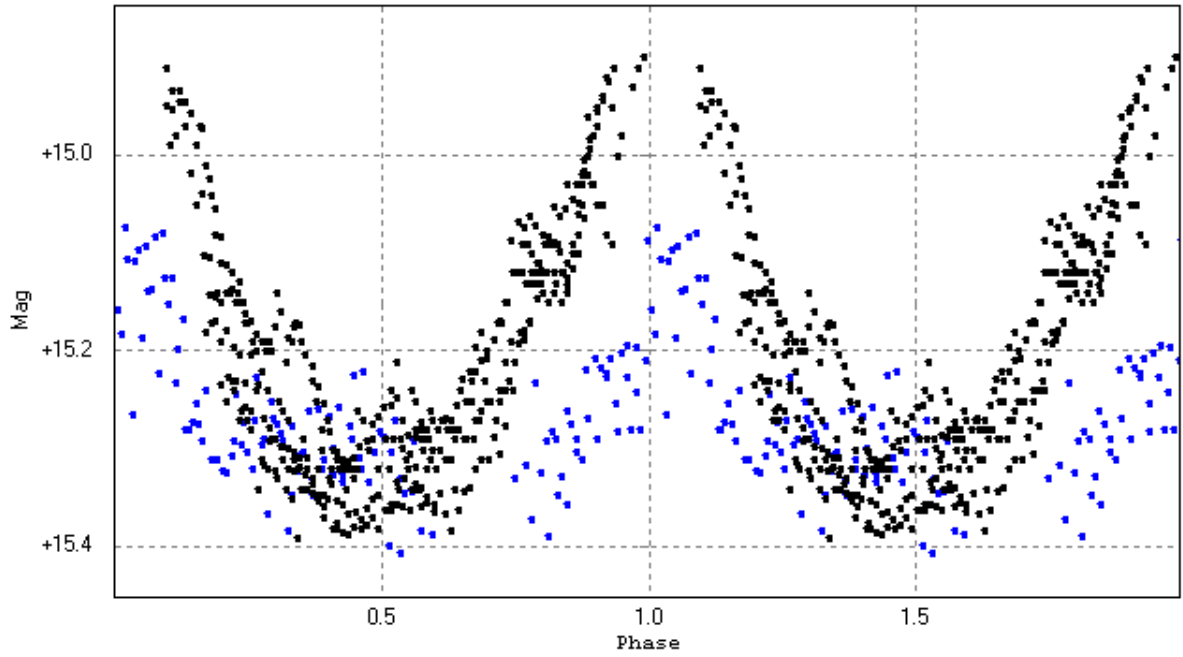

**Figure 6: Phase diagram of data from JD 258 and 263-264 (June 7 and 12-13) folded on $P_{sh} = 0.0653$ d**

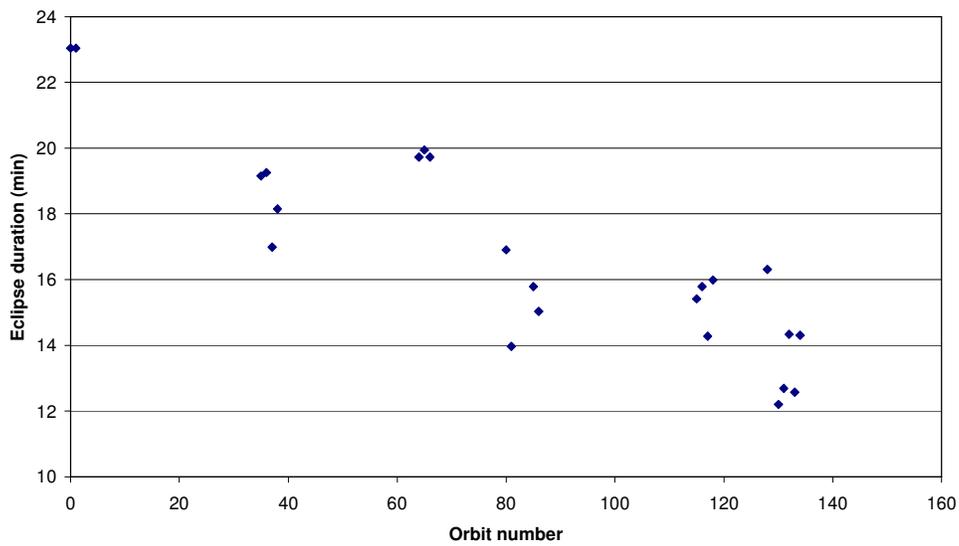

**Figure 7: Duration of eclipses during the outburst**

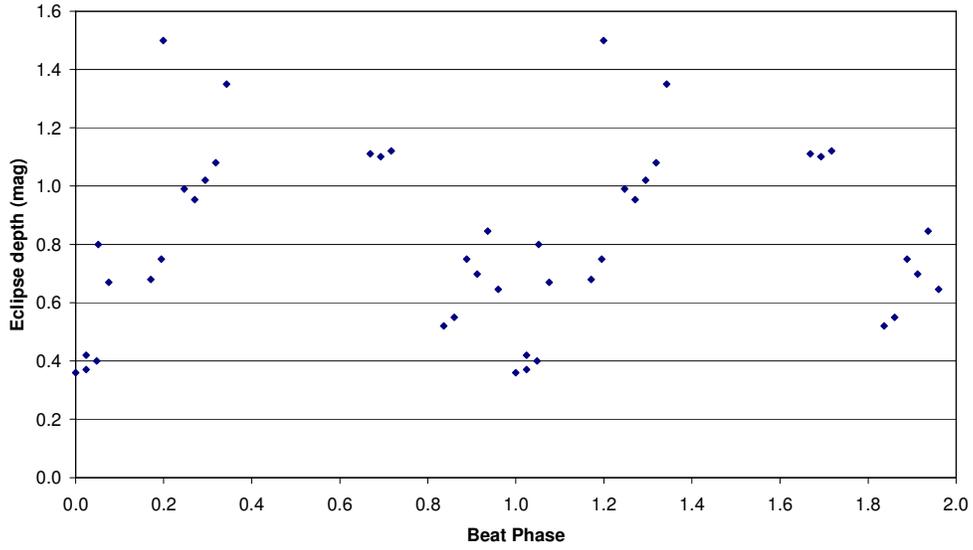

**Figure 8: Correlation of eclipse depth with beat phase**

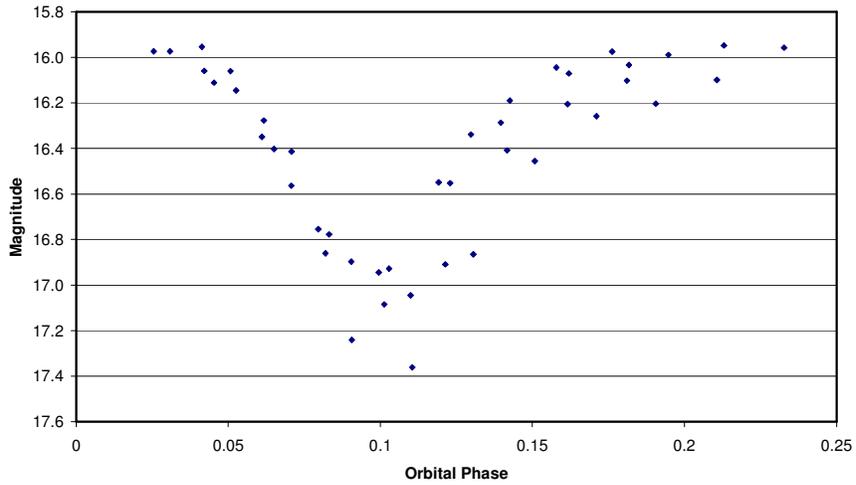

**Figure 9: Average eclipse profile for JD 264 obtained by folding eclipses onto $P_{orb}$**